\documentclass[aps,prl,twocolumn,superscriptaddress,longbibliography]{revtex4-2}
\usepackage{graphicx}
\usepackage{amssymb,amsmath}
\usepackage{bm}
\usepackage{dcolumn}
\usepackage{float}
\usepackage[OT1]{fontenc} 
\usepackage{url}
\usepackage{mathrsfs}
\usepackage{slashed,comment}
\usepackage{color}
\usepackage{verbatim}
\usepackage{enumitem}
\usepackage{soul,physics}
\usepackage[driverfallback=dvipdfm]{hyperref}
\hypersetup{pdfpagemode=FullScreen,colorlinks=true,breaklinks,urlcolor=blue,linkcolor=blue,citecolor=blue}

\usepackage{subfigure}
\usepackage{amssymb}
\usepackage{bm}
\usepackage{graphicx}
\usepackage{amsmath,amssymb}
\usepackage{tikz}
\usepackage{tikz-cd}
\usetikzlibrary{arrows}
\usetikzlibrary{intersections}
\usetikzlibrary{shapes.geometric}
\usetikzlibrary{decorations.pathmorphing, patterns,shapes}
\usetikzlibrary{decorations.markings}

\tikzset{
  mid arrow/.style={postaction={decorate,decoration={
        markings,
        mark=at position .575 with {\arrow[#1]{stealth}}
      }}},
  near arrow/.style={postaction={decorate,decoration={
        markings,
        mark=at position .275 with {\arrow[#1]{stealth}}
      }}},
   far arrow/.style={postaction={decorate,decoration={
        markings,
        mark=at position .800 with {\arrow[#1]{stealth}}
      }}},
}

\begin{document}
  
  \title{Strong-to-weak Symmetry Breaking and Entanglement Transitions }

  \author{Langxuan Chen}
  \affiliation{School of Physics, Xi'an Jiaotong University, Xi'an 710049, China}

  \author{Ning Sun}
  \affiliation{Department of Physics, Fudan University, Shanghai, 200438, China}

  \author{Pengfei Zhang}
  \thanks{PengfeiZhang.physics@gmail.com}
  \affiliation{Department of Physics, Fudan University, Shanghai, 200438, China}
  \affiliation{State Key Laboratory of Surface Physics, Fudan University, Shanghai, 200438, China}
  \affiliation{Shanghai Qi Zhi Institute, AI Tower, Xuhui District, Shanghai 200232, China}
  \affiliation{Hefei National Laboratory, Hefei 230088, China}

  \date{\today}

  \begin{abstract}
  When interacting with an environment, the entanglement within quantum many-body systems is rapidly transferred to the entanglement between the system and the bath. For systems with a large local Hilbert space dimension, this leads to a first-order entanglement transition for the reduced density matrix of the system. On the other hand, recent studies have introduced a new paradigm for classifying density matrices, with particular focus on scenarios where a strongly symmetric density matrix undergoes spontaneous symmetry breaking to a weak symmetry phase. This is typically characterized by a finite R\'enyi-2 correlator or a finite Wightman correlator. In this work, we study the entanglement transition from the perspective of strong-to-weak symmetry breaking, using solvable complex Brownian SYK models. We perform analytical calculations for both the early-time and late-time saddles. The results show that while the R\'enyi-2 correlator indicates a transition from symmetric to symmetry-broken phase, the Wightman correlator becomes finite even in the early-time saddle due to the single-replica limit, demonstrating that the first-order transition occurs between a near-symmetric phase and a deeply symmetry-broken phase in the sense of Wightman correlator. Our results provide a novel viewpoint on the entanglement transition under symmetry constraints and can be readily generalized to systems with repeated measurements.
  \end{abstract}
  
  \maketitle

  \begin{figure}[t]
    \centering
    \includegraphics[width=0.9\linewidth]{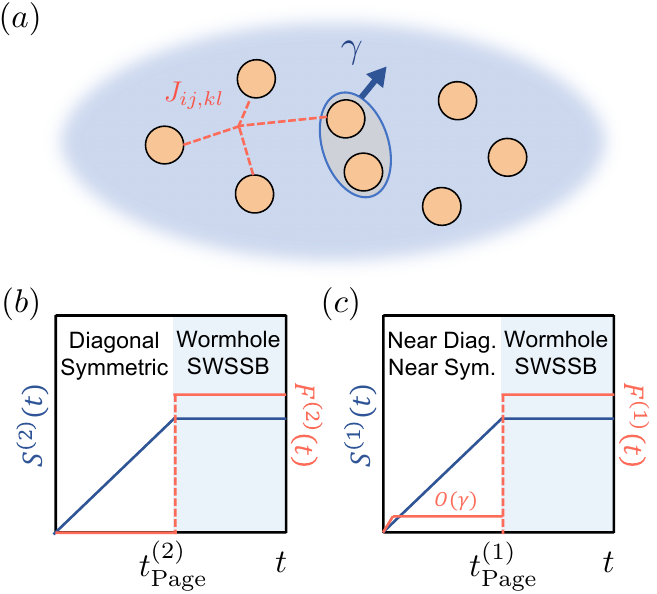}
    \caption{(a). We present a schematic of our model, which consists of a complex Brownian SYK model coupled to an environment. $J_{ij,kl}$ denotes the SYK interaction strength, while $\gamma$ represents a weak system-environment coupling. (b-c) We summarize our main results for (b) the two-replica case and (c) the single-replica limit. The entanglement transition is accompanied by a jump in either the R\'enyi-2 correlator or the Wightman correlator. The R\'enyi-2 correlator vanishes in the early-time saddle, while the Wightman correlator is small but finite due to the replica limit $n\rightarrow 1$. }
    \label{fig:schemticas}
  \end{figure}

  \emph{ \color{blue}Introduction.--} Recent advancements in quantum science and technology have introduced significant challenges in understanding system-environment coupling. A major theoretical development is the discovery of strong-to-weak spontaneous symmetry breaking (SWSSB), where the system's density matrix undergoes a transition from strong to weak symmetry \cite{Lee:2023fsk,Ogunnaike:2023qyh,Lessa:2024gcw,Sala:2024ply,Huang:2024rml,Gu:2024wgc,Kuno:2024aiw,Kuno:2024aiw,Zhang:2024fpf,Zhang:2024gbx,Liu:2024mme,Weinstein:2024fug,Guo:2024ecx}. Early studies identified SWSSB through a finite R\'enyi-2 correlator, derived via the Choi-Jamiolkowski isomorphism \cite{Lee:2023fsk}. Subsequently, the fidelity correlator was introduced as a universal order parameter for SWSSB \cite{Lessa:2024gcw}, defined in the single-replica limit and equivalent to the Wightman (or R\'enyi-1) correlator \cite{Liu:2024mme,Weinstein:2024fug}. SWSSB, as defined by either criterion, indicates that the density matrix is not symmetrically invertible, while a finite fidelity or Wightman correlator further ensures its stability. Additionally, studies have revealed a close relationship between SWSSB and local recoverability \cite{Lessa:2024gcw}, highlighting its significance in the broader context of quantum information science.

  One the other hand, extensive developments have uncovered many scenarios where novel dynamical transitions occur when computing the entanglement entropy. A celebrated example is the black hole information paradox \cite{Hawking:1975vcx}, where recent resolutions highlighted a first-order transition between the replica (almost) diagonal solution at early times and replica wormholes at late times \cite{almheiri2020,almheiri2020a,Almheiri:2019yqk,Penington:2019kki}. Studies \cite{penington2020,chen2020replica,Dadras:2020xfl,jian2021note,Wang:2023vkq} have also observed similar features in quantum many-body systems with large local Hilbert space dimension, such as the Sachdev-Ye-Kitaev (SYK) models \cite{maldacena2016remarks,kitaev2015simple,kitaev2018soft}. A similar transition can also be observed when introducing repeated measurements to higher-dimensional models like SYK chains \cite{jianMeasurementInducedPhaseTransition2021a,zhangUniversalEntanglementTransitions2022,liuNonunitaryDynamicsSachdevYeKitaev2021,zhang2021emergent,zhang2022quantum,PhysRevB.106.224305,Tian-GangZhou:2023jsj}. In this case, the replica diagonal solution becomes area-law entangled, while the replica wormhole solutions exhibit volume-law entanglement. Consequently, the transition between these solutions, as the measurement strength is tuned, corresponds to the measurement-induced entanglement phase transition \cite{mazzucchiQuantumMeasurementinducedDynamics2016,liQuantumZenoEffect2018,skinnerMeasurementInducedPhaseTransitions2019,liMeasurementdrivenEntanglementTransition2019,szyniszewskiEntanglementTransitionVariablestrength2019,chanUnitaryprojectiveEntanglementDynamics2019,vasseurEntanglementTransitionsHolographic2019,zhouEmergentStatisticalMechanics2019,gullansScalableProbesMeasurementInduced2020,jianMeasurementinducedCriticalityRandom2020,fujiMeasurementinducedQuantumCriticality2020,zabaloCriticalPropertiesMeasurementinduced2020,gullansDynamicalPurificationPhase2020,choiQuantumErrorCorrection2020,baoTheoryPhaseTransition2020}. These entanglement transitions reveal a qualitative change in the reduced density matrix of the subregion. Despite this close relationship, a study of entanglement transitions from the perspective of SWSSB remains lacking.

  In this work, we investigate the symmetry patterns of entanglement transitions by computing the R\'enyi-2 correlator and the Wightman correlator in both replica (almost) diagonal and replica wormhole solutions, using a complex Brownian SYK solvable model \cite{Chen:2020bmq,Agarwal:2020yky,Zhang:2023vpm,Cheng:2024fwk}. We assume the Lindblad operator to be charge-neutral, ensuring the strong symmetry of the density matrix, as illustrated in FIG. \ref{fig:schemticas} (a). The results are sketched in FIG. \ref{fig:schemticas} (b-c). The calculation in FIG. \ref{fig:schemticas} (b) is performed with two replicas, and the entanglement transition of the second R\'enyi entropy occurs at $t^{(2)}_{\text{Page}}$. This entanglement transition is accompanied by a jump in R\'enyi-2 correlator from zero to a (near) maximal value, indicating a first-order transition from symmetric phase to deep SWSSB phase in the sense of R\'enyi-2 correlator. However, when considering the limit of a single replica (see FIG. \ref{fig:schemticas} (c)), the early-time saddle also exhibits a small Wightman correlator, demonstrating the manifestation of SWSSB in terms of the Wightman correlator. Nevertheless, its Wightman correlator vanishes linearly with the decoherence strength, and consequently, we refer to it as the near-symmetric phase. 

  \emph{ \color{blue}Setup.--} We consider a system consisting of $N$ complex fermions $c_i$ with $i=1,2,...,N$, interacting via all-to-all Brownian random interactions. This model, known as the Brownian complex SYK model \cite{Chen:2020bmq,Agarwal:2020yky,Zhang:2023vpm,Cheng:2024fwk}, is described by the time-dependent Hamiltonian
  \begin{equation}
  H(t)=\sum_{i<j,k<l}J_{ij,kl}(t)~c_i^\dagger c_j^\dagger c^{}_kc^{}_l.
  \end{equation}
  Here, the canonical commutation relation reads $\{c^{}_i,c_j^{}\}=\delta_{ij}$. The couplings $J_{ij,kl}(t)$ with different indices are independent Brownian variables with 
  \begin{equation}
  \overline{J_{ij,kl}(t)J_{ij,kl}(t')^*}=2J\delta(t-t')/N^{3}.
  \end{equation}
  In addition, the system is coupled to a Markovian environment, and its full evolution is described by the Lindbladian master equation:
  \begin{equation}\label{eqn:mastereqn}
  \frac{d\rho}{dt}=-i[H(t),\rho]+
  \frac{\gamma}{N}\sum_{ij}\left[L_{ij} \rho L_{ij}^\dagger-\frac{1}{2}\{L_{ij}^\dagger L_{ij}, \rho\}\right].
  \end{equation}
  In this work, to be concrete, we focus on a particular choice of Lindblad operators $L_{ij}=c_ic^\dagger_j$ that conserve the charge $Q_c=\sum_i c_i^\dagger c_i^{}$. The factor of $1/N$ is introduced for maintaining a well-defined large-$N$ limit. We would focus on the weak decoherence limit $\gamma\ll J$. The SYK model coupled to Markovian baths has also been analyzed using master equations in \cite{Kawabata:2022osw,PhysRevD.107.106006,Kulkarni:2021gtt,PhysRevResearch.4.L022068}.

  We prepare the system in a pure state at $t=0$. To facilitate an analytical treatment using the path-integral approach, we choose the initial state to be a maximally entangled state $\rho(0)=|\text{EPR}\rangle\langle\text{EPR}|$ between the system fermion $c_i$ and auxiliary fermions $\xi_i$. In the occupation basis, we have $|\text{EPR}\rangle=\otimes_i \frac{1}{\sqrt{2}}(|10\rangle_i+|01\rangle_i)$. This density matrix exhibits a strong symmetry under the total charge $Q=\sum_i (c_i^\dagger c_i^{}+\xi_i^\dagger \xi_i^{}-1)$, satisfying $e^{-i\theta Q}\rho(0)=\rho(0)$. In addition, since the system is at half-filling, it exhibits particle-hole symmetry. The system is then evolved for a time $t$ under the master equation \eqref{eqn:mastereqn}, which acts non-trivially only on the Hilbert space of the original system. This yields the density matrix $\rho(t)$, which is also strongly symmetric with respect to the charge operator $Q$. We then investigate whether this symmetry is spontaneously broken in the thermodynamical limit $N\rightarrow \infty$. 

  We first demonstrate that the strong symmetry cannot be completely broken by showing that the conventional order parameter $C(t)\equiv \text{tr}[\rho(t) c^{}_ic^\dagger_j]$ vanishes for $i\neq j$. The key ingredient is the suppression of sample fluctuations in the large-$N$ limit \cite{kitaev2018soft,Gu:2019jub}. Therefore, we can safely perform the disorder average over the Brownian variables, yielding $C(t)\approx \overline{C(t)}$. After the ensemble average, the system exhibits an additional symmetry, where $c_j\rightarrow -c_j$ and $\xi_j \rightarrow -\xi_j$ for any individual $j$. Consequently, we obtain $C(t)=0$ for arbitrary time $t$. 

  \emph{ \color{blue}Order Parameters for SWSSB.--} The remaining possibility is the strong-to-weak symmetry breaking. We consider both the R\'enyi-2 correlator $F^{(2)}(t)$ and the Wightman correlator $F^{(1)}(t)$. More generally, we introduce the R\'enyi-$n$ correlator
  \begin{equation}\label{eqn:definition}
  F^{(n)}(t)=\frac{\overline{\text{tr}\big[\rho(t)^{\frac{n}{2}}c^{}_ic^\dagger_j\rho(t)^{{\frac{n}{2}}}c^{}_jc^\dagger_i\big]}}{\overline{\text{tr}\big[\rho(t)^{n}\big]}}.
  \end{equation}
   Here, we have performed the disorder average on both the numerator and denominator separately, following a similar approach to the calculation of $C(t)$. After obtaining generic results for arbitrary even $n$, the R\'enyi-2 correlator $F^{(2)}(t)$ and the Wightman correlator $F^{(1)}(t)$ correspond to $n=2$ or analytical continuation to $n\rightarrow 1$, respectively. Similar correlators can be defined for the auxiliary fermions $\xi_i$, which turn out to be equivalent.

  We calculate Eq. \eqref{eqn:definition} using the path-integral approach. We begin by introducing the graphic representation of the density matrix $\rho$
  \begin{equation}
  \rho(t)~=~
  \begin{tikzpicture}[scale=0.8,baseline={([yshift=-3pt]current bounding box.center)}]

\draw[thick,red,mid arrow] (0pt,-10pt)--(70pt,-10pt);

\filldraw (70pt,-17pt) circle (0pt) node[ right]{ $c$};

\filldraw (0pt,-17pt) circle (0pt) node[ left]{ $\xi$};

\draw[thick,blue,mid arrow] (70pt,-24pt)--(0pt,-24pt);

\draw[thick,dotted] (14pt,-11pt)--(14pt,-24pt);
\draw[thick,dotted] (28pt,-11pt)--(28pt,-24pt);
\draw[thick,dotted] (42pt,-11pt)--(42pt,-24pt);
\draw[thick,dotted] (56pt,-11pt)--(56pt,-24pt);

\draw[thick,red] (0pt,-10pt) arc(-90:-180:5pt and 5pt);
\draw[thick,red] (70pt,-10pt) arc(-90:0:5pt and 5pt);

\draw[thick,blue] (0pt,-24pt) arc(90:180:5pt and 5pt);
\draw[thick,blue] (70pt,-24pt) arc(90:0:5pt and 5pt);

\end{tikzpicture}
  \end{equation}
  Here, the red and blue branches are evolved forward and backward, respectively. The right and left endpoints of the contour correspond to the system fermions $c_i$ or auxiliary fermions $\xi_i$. Dotted lines represent the coupling induced by the Lindblad operator. A graphic representation of $F^{(n)}$ then requires introducing $n$ replicas of the density matrix $\rho(t)$. Taking $n=4$ as an illustrative example, we have 
\begin{equation}
  F^{(n)}(t)~=~
  \begin{tikzpicture}[scale=0.8,baseline={([yshift=-3pt]current bounding box.center)}]

\draw[thick,dotted] (14pt,-1pt)--(14pt,-6pt);
\draw[thick,dotted] (28pt,-1pt)--(28pt,-6pt);
\draw[thick,dotted] (42pt,-1pt)--(42pt,-6pt);
\draw[thick,dotted] (56pt,-1pt)--(56pt,-6pt);

\draw[thick,blue] (70pt,-6pt)--(0pt,-6pt);
\draw[thick,blue] (0pt,-6pt) arc(90:180:2pt and 2pt);
\draw[thick,blue] (70pt,-6pt) arc(90:0:2pt and 2pt);

\draw[thick,red] (0pt,-10pt)--(70pt,-10pt);
\draw[thick,red] (0pt,-10pt) arc(-90:-180:2pt and 2pt);
\draw[thick,red] (70pt,-10pt) arc(-90:0:2pt and 2pt);

\draw[thick,dotted] (14pt,-11pt)--(14pt,-20pt);
\draw[thick,dotted] (28pt,-11pt)--(28pt,-20pt);
\draw[thick,dotted] (42pt,-11pt)--(42pt,-20pt);
\draw[thick,dotted] (56pt,-11pt)--(56pt,-20pt);

\draw[thick,blue] (70pt,-20pt)--(0pt,-20pt);
\draw[thick,blue] (0pt,-20pt) arc(90:180:2pt and 2pt);
\draw[thick,blue] (70pt,-20pt) arc(90:0:2pt and 2pt);

\draw[thick,red] (0pt,-24pt)--(70pt,-24pt);
\draw[thick,red] (0pt,-24pt) arc(-90:-180:2pt and 2pt);
\draw[thick,red] (70pt,-24pt) arc(-90:0:2pt and 2pt);

\draw[thick,dotted] (14pt,-25pt)--(14pt,-34pt);
\draw[thick,dotted] (28pt,-25pt)--(28pt,-34pt);
\draw[thick,dotted] (42pt,-25pt)--(42pt,-34pt);
\draw[thick,dotted] (56pt,-25pt)--(56pt,-34pt);

\draw[thick,blue] (70pt,-34pt)--(0pt,-34pt);
\draw[thick,blue] (0pt,-34pt) arc(90:180:2pt and 2pt);
\draw[thick,blue] (70pt,-34pt) arc(90:0:2pt and 2pt);

\draw[thick,red] (0pt,-38pt)--(70pt,-38pt);
\draw[thick,red] (0pt,-38pt) arc(-90:-180:2pt and 2pt);
\draw[thick,red] (70pt,-38pt) arc(-90:0:2pt and 2pt);

\draw[thick,dotted] (14pt,-39pt)--(14pt,-48pt);
\draw[thick,dotted] (28pt,-39pt)--(28pt,-48pt);
\draw[thick,dotted] (42pt,-39pt)--(42pt,-48pt);
\draw[thick,dotted] (56pt,-39pt)--(56pt,-48pt);

\draw[thick,blue] (70pt,-48pt)--(0pt,-48pt);
\draw[thick,blue] (0pt,-48pt) arc(90:180:2pt and 2pt);
\draw[thick,blue] (70pt,-48pt) arc(90:0:2pt and 2pt);

\draw[thick,red] (0pt,-52pt)--(70pt,-52pt);
\draw[thick,red] (0pt,-52pt) arc(-90:-180:2pt and 2pt);
\draw[thick,red] (70pt,-52pt) arc(-90:0:2pt and 2pt);

\draw[thick,dotted] (14pt,-53pt)--(14pt,-58pt);
\draw[thick,dotted] (28pt,-53pt)--(28pt,-58pt);
\draw[thick,dotted] (42pt,-53pt)--(42pt,-58pt);
\draw[thick,dotted] (56pt,-53pt)--(56pt,-58pt);

\filldraw[purple] (70pt,-8pt) circle (1.5pt) node[ right]{$c^{}_ic_j^\dagger$};
\filldraw[purple] (70pt,-36pt) circle (1.5pt) node[ right]{$c^{}_jc_i^\dagger$};

\filldraw (-2pt,-8pt) circle (0pt) node[ left]{\scriptsize$1$};
\filldraw (-2pt,-22pt) circle (0pt) node[ left]{\scriptsize$2$};
\filldraw (-2pt,-36pt) circle (0pt) node[ left]{\scriptsize$3$};
\filldraw (-2pt,-50pt) circle (0pt) node[ left]{\scriptsize$4$};

\end{tikzpicture}
  \end{equation}
We assume periodic boundary conditions along the vertical direction, and the purple circles represent operator insertions. A path-integral representation then requires introducing fermion fields $c_{j,\pm}^{(m)}(u)$ and $\bar{c}_{j,\pm}^{(m)}(u)$ with time variable $u\in[0,t]$. $\pm$ labels forward/backward branches and $m=1,2,...,n$ labels different replicas. The partition function for the path-integral reads
\begin{equation}
\mathcal{Z}=\int Dc_{j,\pm}^{(m)} D\bar{c}_{j,\pm}^{(m)} \exp(-S_0-\delta S_\gamma).
\end{equation}
For later convenience, we separate out the contribution from the system-environment coupling, $\delta S_\gamma$, which takes the explicit form of 
\begin{equation}
\begin{aligned}
\delta S_\gamma&=\frac{\gamma}{2N}\sum_{m,ij}\int du\Big[\bar{L}^{(m)}_{ij,+}(u+\epsilon)L^{(m)}_{ij,+}(u)\\&+\bar{L}^{(m)}_{ij,-}(u)L^{(m)}_{ij,-}(u+\epsilon)-2L^{(m+1)}_{ij,+}(u)\bar{L}^{(m)}_{ij,-}(u)\Big].
\end{aligned}
\end{equation}
Here, $L^{(m)}_{ij,\pm}(u)=c_{i,\pm}^{(m)}(u)\bar{c}_{j,\pm}^{(m)}(u)$ and $\epsilon=0^+$ is introduced for the regularization. 
In the large-$N$ limit, we can take mean fields by performing the Hubbard-Stratonovich transformation. The effective action reads
\begin{equation}\label{eqn:HSaction}
\begin{aligned}
\delta S_\gamma=&\int du~\Bigg[\frac{Nn{\gamma}}{4}+\frac{Nn\phi^2}{\gamma}\\&-\phi\sum_i(\bar{c}_{i,+}^{(m+1)}c_{i,-}^{(m)}+c_{i,+}^{(m+1)}\bar{c}_{i,-}^{(m)})\Bigg].
\end{aligned}
\end{equation}
Here, we have incorporated both particle-hole symmetry and the permutation symmetry between different replicas to reduce the number of independent auxiliary fields. Details are presented in the supplementary material \cite{SM}. $\phi\sim \gamma \bar{c}_{i,+}^{(m+1)}c_{i,-}^{(m)}$ represents the coupling between nearest neighbor replicas. $F^{(n)}(t)$ is then expressed as the four-point function of the path-integral.

  \emph{ \color{blue}Early-time Regime.--} We begin with the calculation in (near) diagonal saddle which dominates the early-time regime. For $\gamma=0$, different replicas are decoupled. It is known that the two-point function decays exponentially in the time difference $G(u)=\langle c_{i,-}^{(m)}(u)\bar{c}_{i,+}^{(m)}(0) \rangle=\frac{1}{2}e^{-\Gamma |u|/2}$, with a quasi-particle decay rate $\Gamma=J/2^{q-2}$ \cite{Chen:2020bmq}. We now add effects of the system-environment coupling perturbatively to determine the saddle-point of $\phi$, assuming the $\phi$ is a constant in time $u$. Two types of contributions appear to the lowest order in $\phi$, represented by graphs (for $n=4$) \cite{Dadras:2019tcz}:
  \begin{equation}
  \delta S_1~=~\begin{tikzpicture}[scale=0.8,baseline={([yshift=-3pt]current bounding box.center)}]

\draw[thick,blue] (70pt,-6pt)--(0pt,-6pt);
\draw[thick,blue] (0pt,-6pt) arc(90:180:2pt and 2pt);
\draw[thick,blue] (70pt,-6pt) arc(90:0:2pt and 2pt);

\draw[thick,red] (0pt,-10pt)--(70pt,-10pt);
\draw[thick,red] (0pt,-10pt) arc(-90:-180:2pt and 2pt);
\draw[thick,red] (70pt,-10pt) arc(-90:0:2pt and 2pt);

\draw[thick,blue] (70pt,-20pt)--(0pt,-20pt);
\draw[thick,blue] (0pt,-20pt) arc(90:180:2pt and 2pt);
\draw[thick,blue] (70pt,-20pt) arc(90:0:2pt and 2pt);

\draw[thick,red] (0pt,-24pt)--(70pt,-24pt);
\draw[thick,red] (0pt,-24pt) arc(-90:-180:2pt and 2pt);
\draw[thick,red] (70pt,-24pt) arc(-90:0:2pt and 2pt);

\draw[thick,blue] (70pt,-34pt)--(0pt,-34pt);
\draw[thick,blue] (0pt,-34pt) arc(90:180:2pt and 2pt);
\draw[thick,blue] (70pt,-34pt) arc(90:0:2pt and 2pt);

\draw[thick,red] (0pt,-38pt)--(70pt,-38pt);
\draw[thick,red] (0pt,-38pt) arc(-90:-180:2pt and 2pt);
\draw[thick,red] (70pt,-38pt) arc(-90:0:2pt and 2pt);

\draw[thick,blue] (70pt,-48pt)--(0pt,-48pt);
\draw[thick,blue] (0pt,-48pt) arc(90:180:2pt and 2pt);
\draw[thick,blue] (70pt,-48pt) arc(90:0:2pt and 2pt);

\draw[thick,red] (0pt,-52pt)--(70pt,-52pt);
\draw[thick,red] (0pt,-52pt) arc(-90:-180:2pt and 2pt);
\draw[thick,red] (70pt,-52pt) arc(-90:0:2pt and 2pt);

\draw[thick,dotted] (28pt,-11pt)--(28pt,-20pt);
\draw[thick,dotted] (42pt,-11pt)--(42pt,-20pt);

\filldraw[purple] (28pt,-10pt) circle (1.2pt) node[ right]{};
\filldraw[purple] (28pt,-20pt) circle (1.2pt) node[ right]{};

\filldraw[purple] (42pt,-10pt) circle (1.2pt) node[ right]{};
\filldraw[purple] (42pt,-20pt) circle (1.2pt) node[ right]{};
\end{tikzpicture}\ \ \ \ \ \ \delta S_2~=~
\begin{tikzpicture}[scale=0.8,baseline={([yshift=-1pt]current bounding box.center)}]

\draw[thick,blue] (70pt,-6pt)--(0pt,-6pt);
\draw[thick,blue] (0pt,-6pt) arc(90:180:2pt and 2pt);
\draw[thick,blue] (70pt,-6pt) arc(90:0:2pt and 2pt);

\draw[thick,red] (0pt,-10pt)--(70pt,-10pt);
\draw[thick,red] (0pt,-10pt) arc(-90:-180:2pt and 2pt);
\draw[thick,red] (70pt,-10pt) arc(-90:0:2pt and 2pt);

\draw[thick,blue] (70pt,-20pt)--(0pt,-20pt);
\draw[thick,blue] (0pt,-20pt) arc(90:180:2pt and 2pt);
\draw[thick,blue] (70pt,-20pt) arc(90:0:2pt and 2pt);

\draw[thick,red] (0pt,-24pt)--(70pt,-24pt);
\draw[thick,red] (0pt,-24pt) arc(-90:-180:2pt and 2pt);
\draw[thick,red] (70pt,-24pt) arc(-90:0:2pt and 2pt);

\draw[thick,blue] (70pt,-34pt)--(0pt,-34pt);
\draw[thick,blue] (0pt,-34pt) arc(90:180:2pt and 2pt);
\draw[thick,blue] (70pt,-34pt) arc(90:0:2pt and 2pt);

\draw[thick,red] (0pt,-38pt)--(70pt,-38pt);
\draw[thick,red] (0pt,-38pt) arc(-90:-180:2pt and 2pt);
\draw[thick,red] (70pt,-38pt) arc(-90:0:2pt and 2pt);

\draw[thick,blue] (70pt,-48pt)--(0pt,-48pt);
\draw[thick,blue] (0pt,-48pt) arc(90:180:2pt and 2pt);
\draw[thick,blue] (70pt,-48pt) arc(90:0:2pt and 2pt);

\draw[thick,red] (0pt,-52pt)--(70pt,-52pt);
\draw[thick,red] (0pt,-52pt) arc(-90:-180:2pt and 2pt);
\draw[thick,red] (70pt,-52pt) arc(-90:0:2pt and 2pt);

\draw[thick,dotted] (28pt,-11pt)--(28pt,-20pt);
\filldraw[purple] (28pt,-10pt) circle (1.2pt) node[ right]{};
\filldraw[purple] (28pt,-20pt) circle (1.2pt) node[ right]{};

\draw[thick,dotted] (42pt,-25pt)--(42pt,-34pt);
\filldraw[purple] (42pt,-24pt) circle (1.2pt) node[ right]{};
\filldraw[purple] (42pt,-34pt) circle (1.2pt) node[ right]{};

\draw[thick,dotted] (33pt,-39pt)--(33pt,-48pt);
\filldraw[purple] (33pt,-38pt) circle (1.2pt) node[ right]{};
\filldraw[purple] (33pt,-48pt) circle (1.2pt) node[ right]{};

\draw[thick,dotted] (37pt,-1pt)--(37pt,-6pt);
\filldraw[purple] (37pt,-6pt) circle (1.2pt) node[ right]{};

\draw[thick,dotted] (37pt,-52pt)--(37pt,-58pt);
\filldraw[purple] (37pt,-52pt) circle (1.2pt) node[ right]{};

\end{tikzpicture}
  \end{equation}
where circles correspond to the insertion of $c_{j,\pm}^{(m)}$ or $\bar{c}_{j,\pm}^{(m)}$. Taking these contributions into account, the results for generic $n$ read
\begin{equation}
\frac{\delta S_\gamma}{t N}=\frac{n\gamma }{4}+\frac{n\phi^2}{\gamma}-\frac{n\phi^2 }{2\Gamma}-\frac{2^{n-1}   \Upsilon \left(n-\frac{1}{2}\right)}{\sqrt{\pi } \Upsilon  (n)} \Gamma^{1-n}{\phi}^n.
\end{equation}
Here, $\Upsilon (n) $ is the Gamma function, and the last two terms correspond to contributions from $\delta S_1$ and $\delta S_2$, respectively. We have neglected terms that are subleading in $(\Gamma t)^{-1}$, with the details provided in the supplementary material \cite{SM}.

We first consider the R\'enyi-2 correlator for $n=2$. Both contributions form $\delta S_1$ and $\delta S_2$ are proportional to $\phi^2 t /\Gamma$. In the limit of small $\gamma$, these contributions are negligible compared to $Nn\phi^2 t/\gamma$. Consequently, the minimization over $\phi$ is achieved at $\phi=0$ and different replicas remain decoupled. Similar results apply to arbitrary $n\neq 2$. Therefore, we find a linear growth second R\'enyi entropy $S^{(2)}=\frac{N\gamma t}{2}$ and
\begin{equation}
F^{(2)}(t)=C(t)^2=0.
\end{equation}
However, an exact replica diagonal solution leads to a divergent R\'enyi entropy $S^{(n)}=\frac{Nn\gamma t}{4(n-1)}$ at $n\rightarrow 1$ and the off-diagonal components are necessary for arbitrary small $\gamma$. In this limit, the contribution $\delta S_2$ dominates over the term $Nn\phi^2 t/\gamma$ at sufficiently small $\phi$. To the leading order in $\gamma/\Gamma$, the minimization leads to
\begin{equation}
\phi=\frac{1}{2} \left(\frac{\gamma}{\sqrt{\pi}}  \Gamma ^{1-n}\frac{\Upsilon \left(n-\frac{1}{2}\right)}{\Upsilon (n)}\right)^{\frac{1}{2-n}} \xrightarrow[]{n\rightarrow 1} \frac{\gamma}{2}.
\end{equation}
This determines the growth of the Von Neumann entropy as $S^{(1)}=s_0 N t$, where $s_0=\frac{\gamma}{2}\ln \left(4\sqrt{e}\frac{\Gamma}{\gamma}\right)$, showing non-analyticity in the decoherence strength $\gamma$ \cite{Dadras:2019tcz}. We further determine the Wightman correlator by
\begin{equation}
F^{(n)}(t)=\left[\begin{tikzpicture}[scale=0.8,baseline={([yshift=-1pt]current bounding box.center)}]

\draw[thick,blue] (70pt,-6pt)--(0pt,-6pt);
\draw[thick,blue] (0pt,-6pt) arc(90:180:2pt and 2pt);
\draw[thick,blue] (70pt,-6pt) arc(90:0:2pt and 2pt);

\draw[thick,red] (0pt,-10pt)--(70pt,-10pt);
\draw[thick,red] (0pt,-10pt) arc(-90:-180:2pt and 2pt);
\draw[thick,red] (70pt,-10pt) arc(-90:0:2pt and 2pt);

\draw[thick,blue] (70pt,-20pt)--(0pt,-20pt);
\draw[thick,blue] (0pt,-20pt) arc(90:180:2pt and 2pt);
\draw[thick,blue] (70pt,-20pt) arc(90:0:2pt and 2pt);

\draw[thick,red] (0pt,-24pt)--(70pt,-24pt);
\draw[thick,red] (0pt,-24pt) arc(-90:-180:2pt and 2pt);
\draw[thick,red] (70pt,-24pt) arc(-90:0:2pt and 2pt);

\draw[thick,blue] (70pt,-34pt)--(0pt,-34pt);
\draw[thick,blue] (0pt,-34pt) arc(90:180:2pt and 2pt);
\draw[thick,blue] (70pt,-34pt) arc(90:0:2pt and 2pt);

\draw[thick,red] (0pt,-38pt)--(70pt,-38pt);
\draw[thick,red] (0pt,-38pt) arc(-90:-180:2pt and 2pt);
\draw[thick,red] (70pt,-38pt) arc(-90:0:2pt and 2pt);

\draw[thick,blue] (70pt,-48pt)--(0pt,-48pt);
\draw[thick,blue] (0pt,-48pt) arc(90:180:2pt and 2pt);
\draw[thick,blue] (70pt,-48pt) arc(90:0:2pt and 2pt);

\draw[thick,red] (0pt,-52pt)--(70pt,-52pt);
\draw[thick,red] (0pt,-52pt) arc(-90:-180:2pt and 2pt);
\draw[thick,red] (70pt,-52pt) arc(-90:0:2pt and 2pt);



\draw[thick,dotted] (53pt,-39pt)--(53pt,-48pt);
\filldraw[purple] (53pt,-38pt) circle (1.2pt) node[ right]{};
\filldraw[purple] (53pt,-48pt) circle (1.2pt) node[ right]{};

\draw[thick,dotted] (57pt,-1pt)--(57pt,-6pt);
\filldraw[purple] (57pt,-6pt) circle (1.2pt) node[ right]{};

\draw[thick,dotted] (57pt,-52pt)--(57pt,-58pt);
\filldraw[purple] (57pt,-52pt) circle (1.2pt) node[ right]{};

\filldraw[purple] (70pt,-8pt) circle (1.5pt) node[ right]{$c^{}$};
\filldraw[purple] (70pt,-36pt) circle (1.5pt) node[ right]{$\bar{c}$};

\end{tikzpicture}+\begin{tikzpicture}[scale=0.8,baseline={([yshift=-3pt]current bounding box.center)}]

\draw[thick,blue] (70pt,-6pt)--(0pt,-6pt);
\draw[thick,blue] (0pt,-6pt) arc(90:180:2pt and 2pt);
\draw[thick,blue] (70pt,-6pt) arc(90:0:2pt and 2pt);

\draw[thick,red] (0pt,-10pt)--(70pt,-10pt);
\draw[thick,red] (0pt,-10pt) arc(-90:-180:2pt and 2pt);
\draw[thick,red] (70pt,-10pt) arc(-90:0:2pt and 2pt);

\draw[thick,blue] (70pt,-20pt)--(0pt,-20pt);
\draw[thick,blue] (0pt,-20pt) arc(90:180:2pt and 2pt);
\draw[thick,blue] (70pt,-20pt) arc(90:0:2pt and 2pt);

\draw[thick,red] (0pt,-24pt)--(70pt,-24pt);
\draw[thick,red] (0pt,-24pt) arc(-90:-180:2pt and 2pt);
\draw[thick,red] (70pt,-24pt) arc(-90:0:2pt and 2pt);

\draw[thick,blue] (70pt,-34pt)--(0pt,-34pt);
\draw[thick,blue] (0pt,-34pt) arc(90:180:2pt and 2pt);
\draw[thick,blue] (70pt,-34pt) arc(90:0:2pt and 2pt);

\draw[thick,red] (0pt,-38pt)--(70pt,-38pt);
\draw[thick,red] (0pt,-38pt) arc(-90:-180:2pt and 2pt);
\draw[thick,red] (70pt,-38pt) arc(-90:0:2pt and 2pt);

\draw[thick,blue] (70pt,-48pt)--(0pt,-48pt);
\draw[thick,blue] (0pt,-48pt) arc(90:180:2pt and 2pt);
\draw[thick,blue] (70pt,-48pt) arc(90:0:2pt and 2pt);

\draw[thick,red] (0pt,-52pt)--(70pt,-52pt);
\draw[thick,red] (0pt,-52pt) arc(-90:-180:2pt and 2pt);
\draw[thick,red] (70pt,-52pt) arc(-90:0:2pt and 2pt);

\draw[thick,dotted] (48pt,-11pt)--(48pt,-20pt);
\filldraw[purple] (48pt,-10pt) circle (1.2pt) node[ right]{};
\filldraw[purple] (48pt,-20pt) circle (1.2pt) node[ right]{};

\draw[thick,dotted] (62pt,-25pt)--(62pt,-34pt);
\filldraw[purple] (62pt,-24pt) circle (1.2pt) node[ right]{};
\filldraw[purple] (62pt,-34pt) circle (1.2pt) node[ right]{};




\filldraw[purple] (70pt,-8pt) circle (1.5pt) node[ right]{$c^{}$};
\filldraw[purple] (70pt,-36pt) circle (1.5pt) node[ right]{$\bar{c}$};

\end{tikzpicture}\right]^2.
\end{equation}
Leaving the details to the supplementary material \cite{SM}, the result (for $t\Gamma \gg 1$) is given by
\begin{equation}\label{eqn:early}
F^{(n)}(t)=\frac{2^n}{\pi}\frac{\Upsilon(\frac{n+1}{2})^2}{\Upsilon(\frac{n}{2}+2)^2}\left(\frac{\phi}{\Gamma}\right)^n \xrightarrow[]{n\rightarrow 1} \frac{16}{9\pi^2}\frac{\gamma}{\Gamma}.
\end{equation}
The result demonstrates a weak SWSSB in the sense of the Wightman correlator, showing a qualitative difference compared to the R\'enyi-2 case. Nevertheless, Wightman correlator vanishes in the limit of $\gamma \rightarrow 0$, and we refer to it as a near-symmetric phase in the sense of the Wightman correlator. We highlight this weak SWSSB also has qualitative consequence. In \cite{Liu:2024mme}, it is established that the modular Hamiltonian of density matrices without SWSSB should exhibit spooky properties, possibly with operator sizes scaling with the system size or an unbounded spectrum. A weak SWSSB relaxes this requirement and makes a regular modular Hamiltonian possible.

We further emphasize that the linear dependence on $\gamma$ in Eq. \eqref{eqn:early} is non-trival, reflecting the non-analytical behavior of $G^W=\text{tr}\big[\sqrt{\rho(t)}c^{}_i\sqrt{\rho(t)}c^\dagger_i\big]\sim \sqrt{\gamma}$. It is also straightforward to generalize our calculation to arbitrary systems where the two-point functions are known and the non-analyticity in both $s_0$ and $G^W$ persists. Therefore, we expect our results to be readily testable experimentally on noisy intermediate-Scale quantum devices \cite{Preskill:2018jim}, using protocols to prepare the thermofield double state of the full density matrix $\rho(t)$ \cite{PhysRevLett.123.220502,Zhu:2019bri,PhysRevA.104.012427,Cottrell:2018ash,Su:2020zgc}.

\emph{ \color{blue}Late-time Regime.--} 
The linear growth of entanglement has to stop when $S^{(n)}$ of the (near) diagonal saddle point exceed the maximally allowed entanglement $2N\ln 2$, as a close analog of the information paradox \cite{Hawking:1975vcx}. This determines the Page time $t^{(2)}_{\text{Page}}=4\ln 2/\gamma$ and $t^{(1)}_{\text{Page}}=2\ln 2/s_0$. New saddle point dominates for the late-time regime with $t>t^{(n)}_{\text{Page}}$, known as the replica wormhole solution \cite{penington2020,chen2020replica,Dadras:2020xfl,jian2021note,Wang:2023vkq}. The solution can be understood as replacing the Lindbladian evolution with a projection onto its steady states, which becomes exact in the long-time limit. Noting that the Lindblad operators satisfy $\hat{L}_{ij}=\hat{L}_{ji}^\dagger$, the steady states are maximally mixed states in arbitrary charge sectors. Graphically, this can be represented by
\begin{equation}
  \rho(t)~\approx ~
  \begin{tikzpicture}[scale=0.8,baseline={([yshift=-3pt]current bounding box.center)}]

\draw[thick,red,mid arrow] (15pt,-14pt)--(55pt,-14pt);

\draw[thick,blue,mid arrow] (55pt,-24pt)--(15pt,-24pt);

\draw[thick,dotted] (25pt,-15pt)--(25pt,-24pt);
\draw[thick,dotted] (35pt,-15pt)--(35pt,-24pt);
\draw[thick,dotted] (45pt,-15pt)--(45pt,-24pt);

\draw[thick,red] (0pt,-14pt) arc(-90:-180:5pt and 5pt);
\draw[thick,red] (70pt,-14pt) arc(-90:0:5pt and 5pt);

\draw[thick,blue] (0pt,-24pt) arc(90:180:5pt and 5pt);
\draw[thick,blue] (70pt,-24pt) arc(90:0:5pt and 5pt);

\draw[thick,black] (0pt,-14pt) arc(90:-90:5pt and 5pt);

\draw[thick,black] (55pt,-14pt) arc(90:-90:5pt and 5pt);

\draw[thick,black] (70pt,-14pt) arc(90:270:5pt and 5pt);
\draw[thick,black] (15pt,-14pt) arc(90:270:5pt and 5pt);

\end{tikzpicture}~=~  \begin{tikzpicture}[scale=0.8,baseline={([yshift=-3pt]current bounding box.center)}]

\draw[thick,black] (0pt,-9pt)--(0pt,-29pt);
\draw[thick,red] (0pt,-9pt)--(0pt,-14pt);
\draw[thick,blue] (0pt,-24pt)--(0pt,-29pt);

\draw[thick,black] (20pt,-9pt)--(20pt,-29pt);
\draw[thick,red] (20pt,-9pt)--(20pt,-14pt);
\draw[thick,blue] (20pt,-24pt)--(20pt,-29pt);

\end{tikzpicture}
  \end{equation}
  Here, the forward/backward evolution in different replicas effectively connect together, and behaves as a standard Keldysh contour. The resulting density matrix is maximally mixed, leading to $S^{(n)}=2N\ln 2$ as expected. In terms of the Hubbard-Stratonovich field $\phi$, this corresponds to having $\langle \bar{c}_{i,+}^{(m+1)}(u)c_{i,-}^{(m)}(u)\rangle =1/2$ and $\phi=\gamma/2$ for arbitrary $n$. We now apply this graphic relation to compute the R\'enyi-$n$ correlator, which becomes
\begin{equation}
 F^{(n)}(t)~=~
  \begin{tikzpicture}[scale=0.8,baseline={([yshift=-3pt]current bounding box.center)}]

\draw[thick] (70pt,-1pt)--(70pt,-58pt);

\draw[thick] (0pt,-1pt)--(0pt,-58pt);

\draw[thick,dotted] (25pt,-1pt)--(25pt,-6pt);
\draw[thick,dotted] (35pt,-1pt)--(35pt,-6pt);
\draw[thick,dotted] (45pt,-1pt)--(45pt,-6pt);

\draw[thick,blue] (55pt,-6pt)--(15pt,-6pt);

\draw[thick,red] (15pt,-10pt)--(55pt,-10pt);

\draw[thick,dotted] (25pt,-11pt)--(25pt,-20pt);
\draw[thick,dotted] (35pt,-11pt)--(35pt,-20pt);
\draw[thick,dotted] (45pt,-11pt)--(45pt,-20pt);

\draw[thick,blue] (55pt,-20pt)--(15pt,-20pt);

\draw[thick,red] (15pt,-24pt)--(55pt,-24pt);

\draw[thick,dotted] (25pt,-25pt)--(25pt,-34pt);
\draw[thick,dotted] (35pt,-25pt)--(35pt,-34pt);
\draw[thick,dotted] (45pt,-25pt)--(45pt,-34pt);

\draw[thick,blue] (55pt,-34pt)--(15pt,-34pt);

\draw[thick,red] (15pt,-38pt)--(55pt,-38pt);

\draw[thick,dotted] (25pt,-39pt)--(25pt,-48pt);
\draw[thick,dotted] (35pt,-39pt)--(35pt,-48pt);
\draw[thick,dotted] (45pt,-39pt)--(45pt,-48pt);

\draw[thick,blue] (55pt,-48pt)--(15pt,-48pt);

\draw[thick,red] (15pt,-52pt)--(55pt,-52pt);

\draw[thick,dotted] (25pt,-53pt)--(25pt,-58pt);
\draw[thick,dotted] (35pt,-53pt)--(35pt,-58pt);
\draw[thick,dotted] (45pt,-53pt)--(45pt,-58pt);

\filldraw[purple] (70pt,-8pt) circle (1.5pt) node[ right]{ };
\filldraw[purple] (70pt,-36pt) circle (1.5pt) node[ right]{ };

\draw[thick,black] (15pt,-6pt) arc(-90:-180:5pt and 5pt);
\draw[thick,black] (55pt,-6pt) arc(-90:0:5pt and 5pt);

\draw[thick,black] (15pt,-10pt) arc(90:270:5pt and 5pt);
\draw[thick,black] (55pt,-10pt) arc(90:-90:5pt and 5pt);

\draw[thick,black] (15pt,-24pt) arc(90:270:5pt and 5pt);
\draw[thick,black] (55pt,-24pt) arc(90:-90:5pt and 5pt);

\draw[thick,black] (15pt,-38pt) arc(90:270:5pt and 5pt);
\draw[thick,black] (55pt,-38pt) arc(90:-90:5pt and 5pt);

\draw[thick,black] (15pt,-52pt) arc(90:180:5pt and 5pt);
\draw[thick,black] (55pt,-52pt) arc(90:0:5pt and 5pt);

\end{tikzpicture}=2^{-2N}\text{tr}[c^{}_ic_j^\dagger c^{}_jc_i^\dagger]=\frac{1}{4},
\end{equation}
which is finite for arbitrary $n$. Importantly, it does not depend on the decoherence strength $\gamma$, and the state lies in a deeply SWSSB phase. This demonstrates a large jump of both R\'enyi-2 correlator and Wightman correlator at the Page time.

 \emph{ \color{blue}Discussions.--} 
  In this work, we investigate the dynamical evolution of an all-to-all interacting fermion model coupled to a Markovian bath with a charge-conserving Lindblad operator. By choosing an appropriate initial state, the density matrix of the system exhibits strong symmetry. Through analytical calculations, we demonstrate that the entanglement transition at the Page time can be diagnosed by a change in symmetry patterns. In terms of the R\'enyi-2 correlator, the transition is from a symmetric phase to a deep SWSSB phase, while in terms of the Wightman correlator, it is from a near-symmetric phase, or a phase with weak SWSSB, to a deep SWSSB phase. While our main analysis is based on the complex Brownian SYK solvable model, we expect the qualitative features to hold for other systems with large local Hilbert space dimensions. Consequently, our predictions can be tested experimentally on quantum devices with all-to-all connectivities, such as reconfigurable Rydberg arrays \cite{Evered:2023wbk,Ma:2023ltx,Bluvstein:2023zmt,2019arXiv190407369B,doi:10.1126/science.abg2530,Ebadi:2022oxd,Bluvstein:2021jsq,Lis:2023gaz,Manetsch:2024lwl,Tao:2023uxy,Cao:2024dhc}.

   Our results open up several intriguing directions for further research. First, it is interesting to study higher-dimensional systems, which would enable the exploration of charge transport based on the close relation between charge diffusion and SWSSB \cite{Ogunnaike:2023qyh, Huang:2024rml, Gu:2024wgc}. Second, recent studies \cite{Zhang2024} have pointed out that charge-conserving Lindblad operators prevent the emergence of a dissipative phase from the perspective of operator size growth. Whether this phenomenon can be understood directly from a symmetry perspective warrants further investigation. Finally, entanglement transitions also appear in systems with repeated measurements. It is natural to expect that similar physics will emerge in volume-law or area-law entangled phases, analogous to the late-time or early-time regime. We leave these questions to future work.

 \vspace{5pt}
  \textit{Acknowledgments.}
  We thank Hanteng Wang, and Tian-Gang Zhou for helpful discussions. This project is supported by the NSFC under grant number 12374477 and by Innovation Program for Quantum Science and Technology under grant number 2024ZD0300101.

  \textit{Note added.}
  While finalizing our manuscript, we noticed a recent study on strong-to-weak parity symmetry breaking in Majorana static SYK models \cite{Kim:2024bou}. Their work focuses on the error threshold of the ground state manifold, rather than the entanglement dynamics.

\bibliography{SWSSB_Wormhole.bbl}

\end{document}